\documentclass[a4paper,fleqn,usenatbib]{mnras}
\usepackage{newtxtext,newtxmath}
\usepackage[T1]{fontenc}
\usepackage{ae,aecompl}
\usepackage{graphics}
\usepackage{graphicx}
\usepackage{amsmath}	
\usepackage{booktabs,caption}

\title[1A 1744-361]{Spectral study of neutron star low mass X-ray binary source 1A 1744-361}

\author[Tobrej et. al.]{
Mohammed Tobrej,$^{1}$\thanks{tabrez.md565@gmail.com}
Binay Rai,$^{1}$\thanks{binayrai21@gmail.com}
Manoj Ghising,$^{1}$\thanks{manojghising26@gmail.com}
Ruchi Tamang,$^{1}$\thanks{ruchitamang76@gmail.com}
\newauthor
Bikash Chandra Paul$^{1}$\thanks{bcpaul@associates.iucaa.in}
\\
$^{1}$Department of Physics, North Bengal University, Siliguri, Darjeeling, WB, 734013, India
\\
}

\pubyear{2021}

\begin{document}
\label{firstpage}
\pagerange{\pageref{firstpage}--\pageref{lastpage}}
\maketitle

\begin{abstract}
We present X-ray observations of the recent outburst of 2022 from the neutron star low mass X-ray binary (LMXB) source 1A 1744-361. Spectral properties of the source have been analyzed using joint NuSTAR and NICER observations. During our observations, the source happens to be in the banana state (soft state) of the hardness intensity diagram (HID). In addition to a power-law with a high energy cutoff, the spectrum is found to exhibit broad iron $K_{\alpha}$ emission along with distinct absorption features. A prominent absorption feature observed at 6.92 keV may be interpreted as $K_{\alpha}$ absorption line from hydrogen-like iron. The absorption feature observed at 7.98 keV may be interpreted as a blend of Fe XXV and Ni XXVII transitions. We have summarized the evidence of variability of the spectral features observed in the X-ray continuum by time-resolved spectroscopy.
\end{abstract}


\section{Introduction}
A Low Mass X-Ray Binary (LMXB) is a binary star system in which one of the stars is either a black hole or neutron star. In most cases, the companion star, a donor, fills its Roche lobe and contributes mass to the compact star. In LMXBs harbouring neutron stars, a weakly magnetized neutron star (NS) accretes matter from a low-mass companion star via Roche-lobe overflow. The X-ray characteristics of LMXBs reveal strong variability over timescales (milliseconds to decades). A major sub classification of NS-LMXBs is based on the spectral and timing properties of the sources \citep{37}. LMXBs with weakly magnetized neutron stars have been divided into two
categories, referred to as Z and atoll sources depending
on the trends that they exhibit in the colour-color diagram (CCD) and the hardness-intensity diagram (HID). In the CCD and HID, Z sources trace a 'Z'-shaped pattern while atoll sources produce a C-type pattern. Further, there exists two divisions of the C-type track of atoll sources: island state and banana state. Low count rates and hard energy spectra define the island state while the banana state is characterized by higher luminosities and softer X-ray spectra \citep{Barret}. The recurrent transient source 1A 1744-361 belongs to the class of LMXBs. 1A 1744-361 was discovered by ArielV in 1976, during an outburst \citep{1,2}. This source is an accreting neutron star with detected type I bursts and a 530 Hz burst oscillation \citep{34}.
 
Type I bursts from an X-ray source help in the categorization of the source, as these types of bursts, are a result of thermonuclear burning of matter piled up on the surfaces of accreting neutron stars \citep{13, 14}. Thermonuclear bursts have been known to be observed in LMXB systems \citep{18, 16}. A thermonuclear X-ray burst from the 2005 RXTE-PCA data of this source was detected by \cite{3}, confirming the idea of \cite{4} that the source under consideration hosts a neutron star. This burst revealed millisecond-period brightness oscillations that estimated the spin frequency of the neutron star to be 530 Hz \citep{3}. Such oscillations originate due to an asymmetric brightness pattern on the stellar surface  modulated by the rotation of the star \citep{17, 18}. Energy-dependent dips were observed in the 2003 PCA data, revealing this source as a dipping LMXB (dipper). The orbital period of the source was estimated to be 97 minutes \citep{3}. These types of LMXBs exhibit modulation of soft X-ray intensity with the orbital period. Such modulation is a consequence of structures above the accretion disk \citep{5} which is possible only if the dippers are inclined such that the line of sight passes through these structures. Hence, dippers serve as a source to explore the properties of the upper layers of accretion disks and the photoionized plasma above them \citep{6} in LMXBs. Also, spectral lines have been reported from several dippers, such as EXO 0748-67 \citep{7}, XB 1916-053 \citep{Boirin}, X1624-490 \citep{9}, and XB 1254-690 \citep{Iaria}. Other dippers include X 1624-490 \citep{39} and XB 1916-0539 \citep{Iariaaa}. The eclipsing source MXB 1659-298 is also associated with Fe-K absorption lines \citep{42}. Hence, 1A 1744-361 is an interesting source for understanding such features which could be of immense significance for exploring various X-ray emitting and absorbing components of LMXBs.

Post discovery, the source has been reported to undergo outbursts in 1989, 2003, 2004, 2005, 2013, and 2022. The outbursts were detected by RXTE (ASM and PCA), Chandra, and INTEGRAL observations \citep{10, 11, 12, 19, 20, 21}. The most recent outburst activity of the source 1A 1744-361 was observed by various observatories. The outburst reported by MAXI/GSC detected from the field including 1A 1744-361 \citep{22}  was subsequently confirmed by Swift/XRT \citep{23} observations. Follow-up observations made by NICER \citep{24} and NuSTAR \citep{25} have been used to interpret the properties of the source. Pulsation searches over the averaged power spectra do not reveal any significant coherent periodicity. Although type-I X-ray bursts were previously observed from the source, we do not observe any type-I X-ray bursts in the NuSTAR light-curve in the present study.  
 
In the paper, we present the X-ray observations of the source using NuSTAR and NICER observations. Spectral analysis of the source reveals typical continuum features, with additional absorption lines from highly ionized Fe in gas that is along the line of sight, above the accretion disk. We report the presence of prominent absorption lines alongside interesting reflection/ emission features in the X-ray spectra of the source.

\section{Observation and Data reduction}
The source 1A 1744-361 was reported to encounter an outburst recently and it was observed by various observatories. For our analysis, we have considered observations from Nuclear Spectroscopic Telescope Array (NuSTAR) and Neutron Star Interior Composition Explorer (NICER).

\subsection{NuSTAR}
The source was observed by NuSTAR observatory on the 08 th of June, 2022. The data reduction has been carried out by  HEASOFT v6.30.1 \footnote{\url{https://heasarc.gsfc.nasa.gov/docs/software/heasoft/download.html}} and CALDB v20220525. NuSTAR is sensitive in the (3-79) keV broadband energy range. It consists of two similar but unidentical co-aligned X-ray focusing telescopes which focus X-rays onto two independent Focal Plane Modules - FPMA and FPMB consisting of a pixelated solid state CdZnTe detector \citep{27}. The mission-specific NUPIPELINE was used for obtaining the clean event files. The XSELECT tool was used for reading the obtained clean event files for extracting the required light curves, spectra and images. The image was observed using astronomical imaging and data visualization application DS9\footnote{\url{https://sites.google.com/cfa.harvard.edu/saoimageds9}}. A circular region of 70 arcsec
is considered as the source extraction region in the analysis. Another region of the same radius in the source free region is considered as the background region. The mission-specific tool NUPRODUCTS has been used to extract the lightcurves, spectra, response matrix files (RMFs), and the corresponding ancillary response files (ARFs). The required background correction for the light curve was done using LCMATH task of FTOOLS. The NuSTAR data has been barycentered to the solar system barycenter using the FTOOL BARYCORR. The obtained spectrum was fitted in XSPEC version 12.12.0 \citep{28}. The specifications regarding the NuSTAR observation has been given in Table 1.
\subsection{NICER}
NICER is an International Space Station (IIS) external payload that has been designed for studying neutron star systems via soft X-ray Timing \citep{29}. It comprises of 56 aligned Focal Plane Modules (FPMs) from which 52 are currently operational. The two detectors (14,34) are associated with an increase in detector noise. Each FPM is comprised of a detector, a preamplifier, and a thermoelectric cooler. Here, we have used X-ray Timing Instrument (XTI) sensitive in the (0.2-12.0) keV energy range. The standard data screening and reduction of NICER observations have been carried out by NICERDAS v9 and CALDBv XTI20210720. The background estimation corresponding to each observation has been made using the tool nibackgen3C50 v6 \citep{30}. The clean event files were filtered using NICERL2. The filtered event files were loaded into XSELECT to extract the required source lightcurve, X-ray spectrum, and response files. The spectra in the energy range (0.7-10.0) keV have been considered for fitting. We have neglected the spectra below 0.7 keV to counter the presence of low-energy noises. Also, we skipped the  spectra above 10 keV  due to background contamination.
 
\begin{table*}
\begin{center}
\begin{tabular}{clllllc}
\hline
Observatory	& Observation ID	& Time of Observation (MJD)	& Exposure (s)  & Mean Count Rate (c/s) \\	
\hline
NuSTAR	& 90801312001 &59738.83 (19:36:09)	&34260  &78.47 $\pm0.08$  \\
NICER   & 5202800106  &59738.19 (04:38:00) &998    &865.7 $\pm1.1$	\\
\hline
\end{tabular}

\caption{NuSTAR and NICER observations represented by the observation ID along with the date of observation, exposure, and mean count rate.}  

\end{center}
\end{table*} 
\begin{figure}

\begin{center}
\includegraphics[angle=0,scale=0.3]{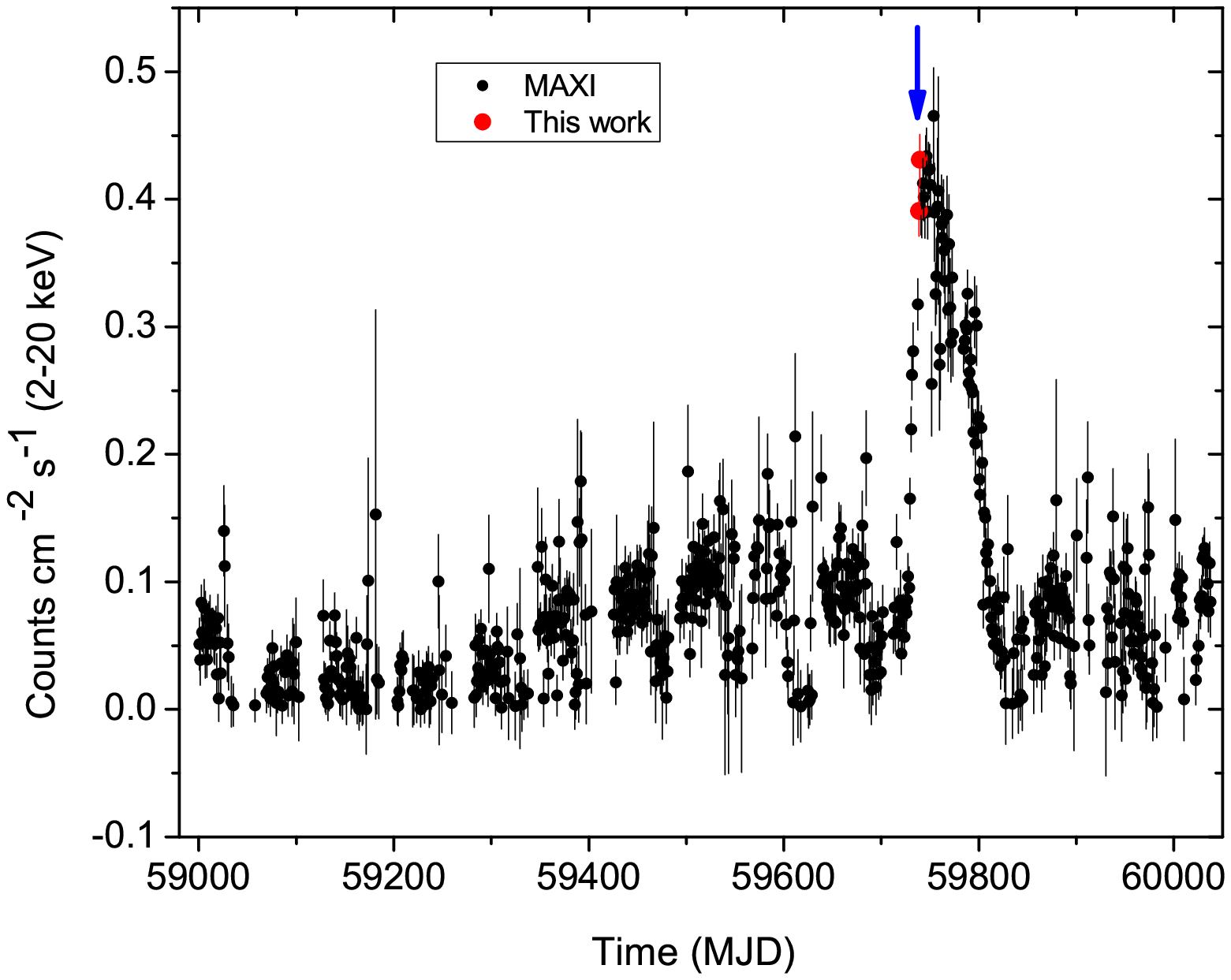}
\end{center}
\caption{MAXI light curve of 1A 1744-361 during its 2022 X-ray outburst in the (2-20) keV energy band. The observations corresponding to the present work are highlighted in red and shown with a vertical arrow.}
\label{1}
\end{figure}

\begin{figure}

\begin{center}
\includegraphics[angle=270,scale=0.3]{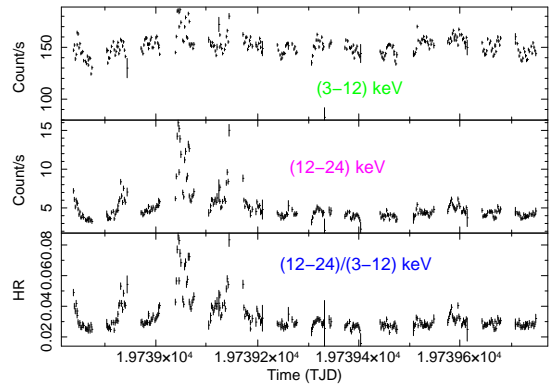}
\end{center}
\caption{Background subtracted light curves of 1A 1744-361 obtained from NuSTAR in the two energy bands of (3-12) keV and (12-24) keV and the Hardness Ratio (HR) variation with time.}
\label{2}
\end{figure}
\section{Lightcurves, hardness-intensity diagram (HID) and color-color diagram (CCD)}

The MAXI Lightcurve of the source in the (2-20) keV energy band is presented in Figure \ref{1}. The observations used in this study are positioned almost at the peak of the outburst. The variability of the light curve provides insights into the physical conditions and the dynamic processes taking place in the X-ray source. For instance, the light curve may exhibit periodic variations, indicating the presence of a compact object, such as a neutron star or black hole, in a binary system. The spectral properties of such a system can vary in a periodic manner, corresponding to the orbital motion of the compact object and the accretion of matter from its companion. To investigate the hardness ratio (HR) relative to time, we have considered background subtracted NuSTAR light curves in two energy bands, (3-12) keV and (12-24) keV (Figure \ref{2}). The HR  has been defined as the ratio of background subtracted light curves in (12-24) keV and (3-12) keV energy ranges. The HR is found to evolve significantly relative to time with a maximum spectral hardening $\sim0.08$. 

\begin{figure}

\begin{center}
\includegraphics[angle=0,scale=0.3]{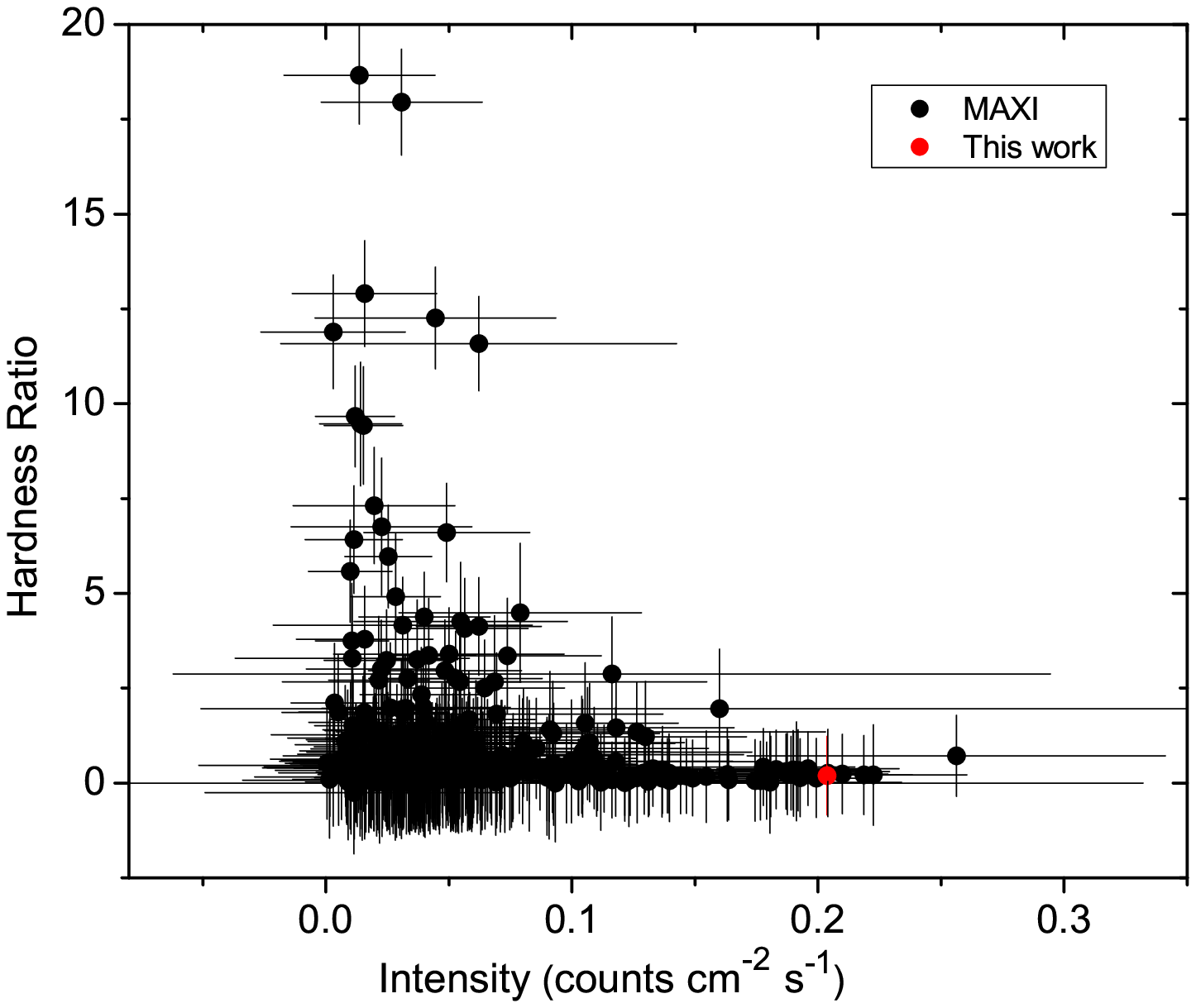}
\end{center}
\caption{Hardness intensity diagram (HID) of 1A 1744-361 using MAXI data.}
\label{3}
\end{figure}

The MAXI Lightcurve of the source  in the (2-20) keV energy band is presented in Figure \ref{1}. The observations used in this study are positioned almost at the peak of the outburst.The hardness-intensity diagram (HID) is produced using MAXI data in the energy bands (10-20) and (4-10) keV. The HID is presented in Figure \ref{3}. The ratio of count rates in the (10-20) keV and (4-10) keV energy bands is considered as the hardness ratio and the overall sum of count rates in the two energy bands is the intensity. It is observed that high intensities correspond to the relatively flat hardness of the banana state whereas the low-intensity data demonstrate the hardening of the island state. The source happens to be in the soft state (banana state) of the hardness intensity diagram (HID) during our observations.
 
\begin{figure}

\begin{center}
\includegraphics[angle=0,scale=0.3]{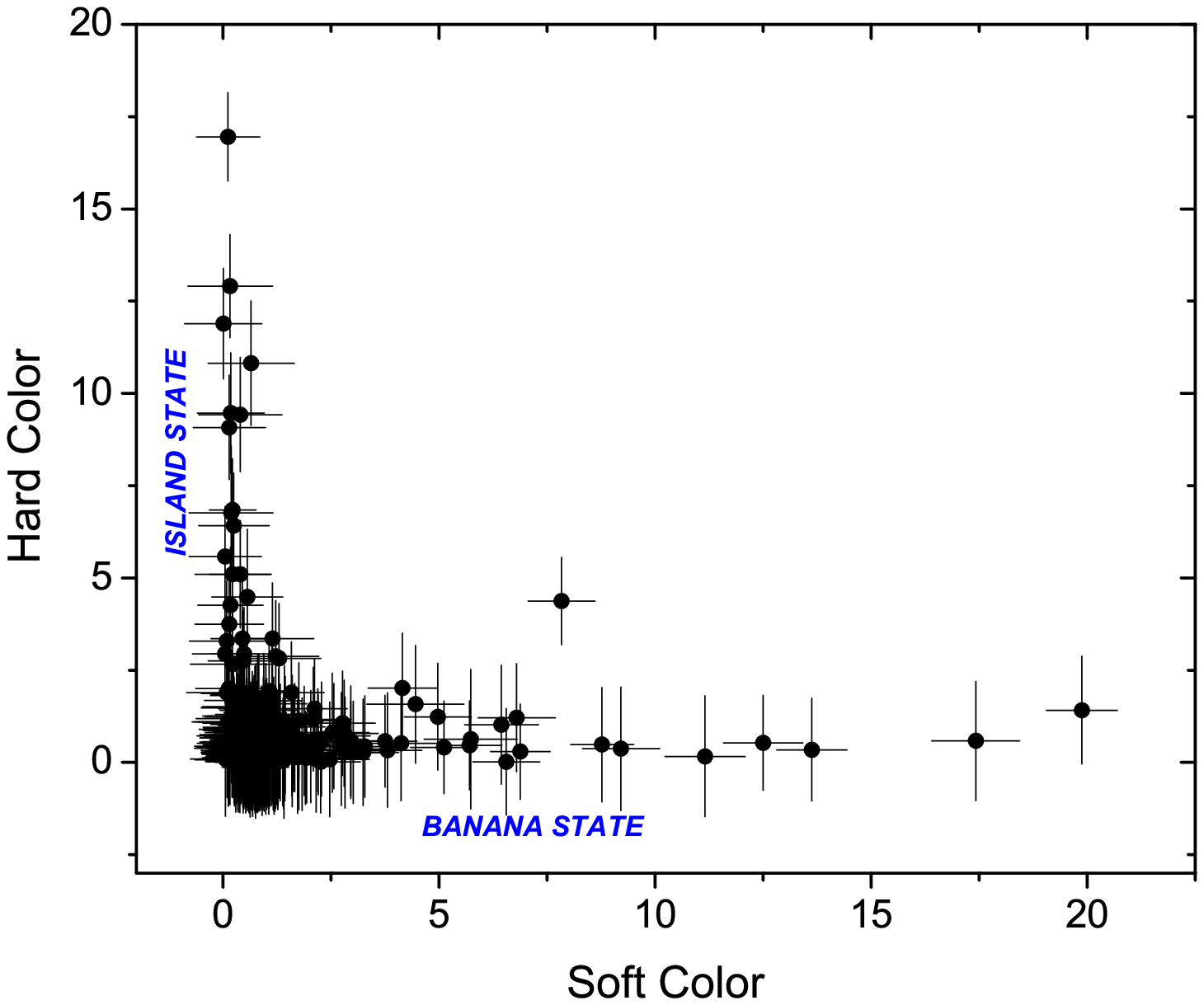}
\end{center}
\caption{CCD of the atoll source 1A 1744-361 obtained using using MAXI data in the period (59000.5-60037.5) MJD. The island and banana spectral states are observed.}
\label{4}
\end{figure}

A color-color diagram (CCD) has been produced by defining the soft colour as the ratio of count rates in the bands 4-10 keV and 2-4  keV, and the hard colour as the ratio of count rates in the bands 10-20 keV and 4-10 keV respectively. The resulting CCD is presented in Figure \ref{4}. The soft-state observations correlate to the banana state, whereas the hard-state observations represent the island state.


 \section{Spectral Analysis}
\begin{figure}

\begin{center}
\includegraphics[angle=270,scale=0.31]{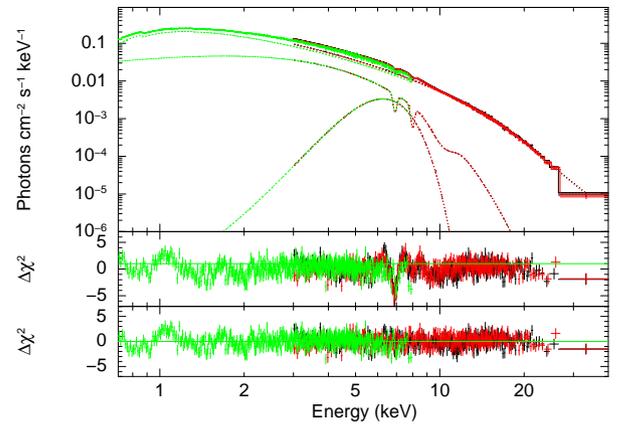}
\end{center}
\caption{The joint spectra corresponding to NuSTAR/NICER observation of 1A 1744-361 in (0.7-40) keV energy range. The top panel represents the unfolded spectra and the middle panel represents the residuals without the incorporation of GABS components. The bottom panel represents the residuals after including the GABS components. Black and red colours represent NuSTAR FPMA and FPMB spectra respectively while green colour represents the NICER spectra.}
\label{5}
\end{figure}

In this section, we present a comprehensive spectral analysis of the source during its recent outburst of 2022. The analysis is based on the spectral fitting of simultaneous NuSTAR and NICER observations carried out on 59738 MJD. GRPPHA tool was used to group the data in such a way that a minimum of 25 counts per spectral bin is obtained. A CONSTANT model was used for simultaneous fitting of NICER and NuSTAR spectra to account for the instrumental uncertainties and non-simultaneity of the observations. We have taken care of the relative normalization factors between the two NuSTAR modules by freezing the constant factor corresponding to instrument FPMA as unity without setting any restriction on FPMB to maintain comparable average count rates in the two instruments. The constant factor corresponding to instrument FPMB  was found to be 0.98 $\;\pm\;0.001$ revealing an uncertainty of $\sim 2\% $ which is in accordance with \cite{32}. We have performed the
spectral fitting of the source using two continuum models. Firstly, the best fit results are obtained using a power law model with a high energy cut-off  along with a soft thermal component (BB). Secondly, we have replaced the CUTOFFPL with the COMPTT model which describes the Comptonization of the soft photon in hot plasma. We denote the two continuum models as model I and model II respectively. The absorption column density has been modelled using the TBABS component with abundance from \cite{33}. The cross-section for the TBABS component was taken as vern \citep{Vern}. Initially, the spectral fitting was carried out using the continuum model (CONSTANT*TBABS*CUTOFFPL). At this point, the spectral fit was found to resemble unacceptable statistics. The quotient spectra obtained using this continuum model was found to resemble distinct negative and positive residuals.  An additional soft thermal component with kT $\sim 1.11$ keV is required to describe the spectrum. The soft excess observed in the spectra was fitted by incorporating a BLACK-BODY (BB) model. At this stage, the fit statistics ($\chi^{2}$ per degrees of freedom) was found to be (3634.66/1853) $\sim 1.96$. Thereafter, we incorporated a Gaussian component to fit the broad iron emission line which led to a significant improvement in the spectral fit. The centroid energy of the feature is estimated to be $\sim 6.37$ keV with an equivalent width of $\sim 1$ keV. The fit statistics after including the GAUSSIAN model was found to be (2876.04/1842) $\sim 1.56$. The absorption column (nH) is estimated to be $0.53\times 10^{22} cm^{-2}$. In addition to the broad Fe emission line, the X-ray spectra was found to exhibit prominent absorption features which have been fitted using the GAUSSIAN absorption component (GABS).  A deep trough observed at $\sim 6.92$ keV, indicates the presence of absorption by highly ionized iron. As evident from Figure \ref{5}, the absorption feature  was fitted by a GABS model which improved the fit statistics to (2592.17/1839) $\sim 1.41$. The equivalent width of the absorption line was found to be 0.12 keV. Another prominent absorption feature at $\sim 7.98$ keV was observed in the X-ray spectra which has been fitted by the inclusion of another GABS component which improved the fit statistics to (2368.51/1836) $\sim1.29$. Furthermore, the X-ray continuum was  revealed distinct negative residuals at $\sim (10-11)$ keV that was fitted by the inclusion of another GABS component. It led to a significant improvement in the fit statistics. The equivalent width of this feature observed at $\sim 10.24$ keV is 0.93 keV. Constraining the strength corresponding to this feature was difficult, leading to insensitivity in the spectral fit results. Therefore, we fixed it at 2 keV after cross-verifying that all the other spectral parameters remain unaffected. The physical significance of this feature is unexplained and may be associated with an instrumental origin, as NuSTAR responses reveal higher uncertainties around the tungsten edge at $\sim$ 10 keV \citep{32}. Finally, the applied continuum model estimated the best fit statistics to be (2169/1833) $\sim 1.18$. The power-law photon index is found to be $\sim 1.14$ with the cut-off energy estimated as $\sim 3.81$ keV. The absorption line observed at $\sim6.92$ keV may be interpreted as $K_\alpha$ line from hydrogen-like iron (e.g., \cite {Neilsen, Honghui Liu et al 2022 ApJ 933 122, Ueda}). The absorption feature at $\sim7.98$ KeV may be attributed to a blend of Fe XXV and Ni XXVII transitions \citep{Neilsen}. The best fit parameters have been presented in Table 3. The absorbed flux in the (0.7-40) keV energy range has been estimated to be $\sim3.71 \;\times\;10^{-9}\;erg\;cm^{-2}\;s^{-1}$ which resembles an X-ray luminosity of $\sim\;3.59 \;\times\;10^{37} erg\;s^{-1}$ assuming the distance to the source as 9 kpc \citep{34}. The spectral fits obtained using models I and II are found to be in agreement with one another. The spectral fit results obtained using the two continuum models do not reveal any appreciable deviations. The NuSTAR observations of \citep{25,Pike} revealed the presence of
two narrow absorption lines that are fairly consistent with
our observations. The spectra in the broad-band (0.7-40) keV energy range has been presented in Figure \ref{5}.

The significance of the absorption lines has been
investigated by simulating the spectra using the \textit{ftest}
command in XSPEC. The \textit{ftest} probabilty is found to
be 2 $\times10^{-10}$ suggesting the presence of absorption lines at $\sim 6.92\; keV, 7.98\; keV, and 10.24\; keV$ respectively. So, the probabilistic origin of these absorption lines can be disregarded. There are no absorption features associated with the NUSTAR and NICER detectors at around
(6-8) keV. Therefore, an instrumental origin of the absorption lines observed at 6.92 keV, and 7.98 keV can be safely ruled out. However, at around 10 keV, the tungsten (W) L-edge is known to reveal residuals \citep{32}. So, it is highly probable that the absorption feature may be associated with an instrumental origin. 
  
\subsection{TIME RESOLVED SPECTROSCOPY}
\begin{figure}

\begin{center}
\includegraphics[angle=0,scale=0.3]{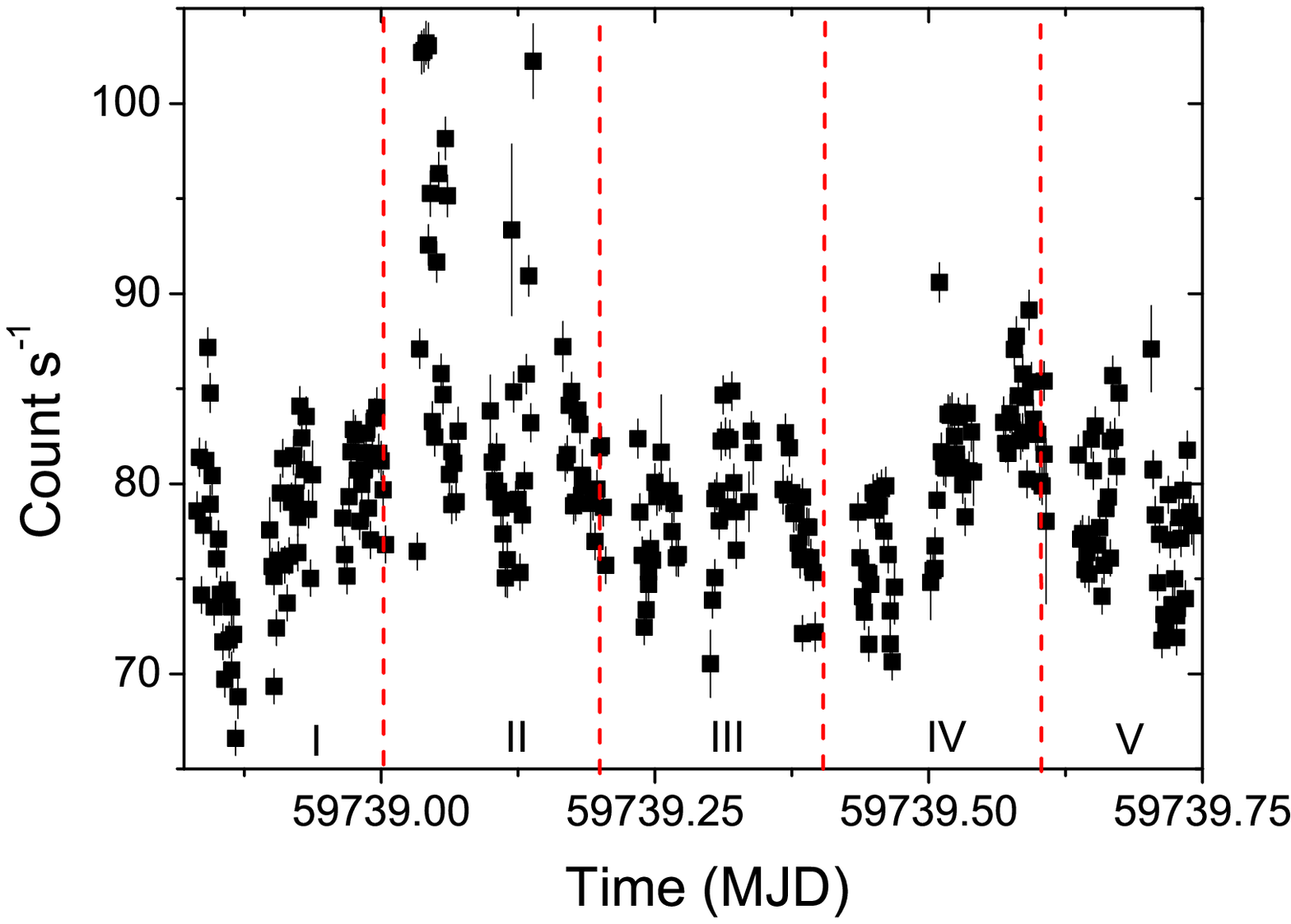}
\end{center}
\caption{NuSTAR light curve of 1A 1744-361 during its 2022 X-ray outburst resolved into five segments. The individual segments are separated by red dashed lines.}
\label{6}
\end{figure}

\begin{figure}

\begin{center}
\includegraphics[angle=0,scale=0.33]{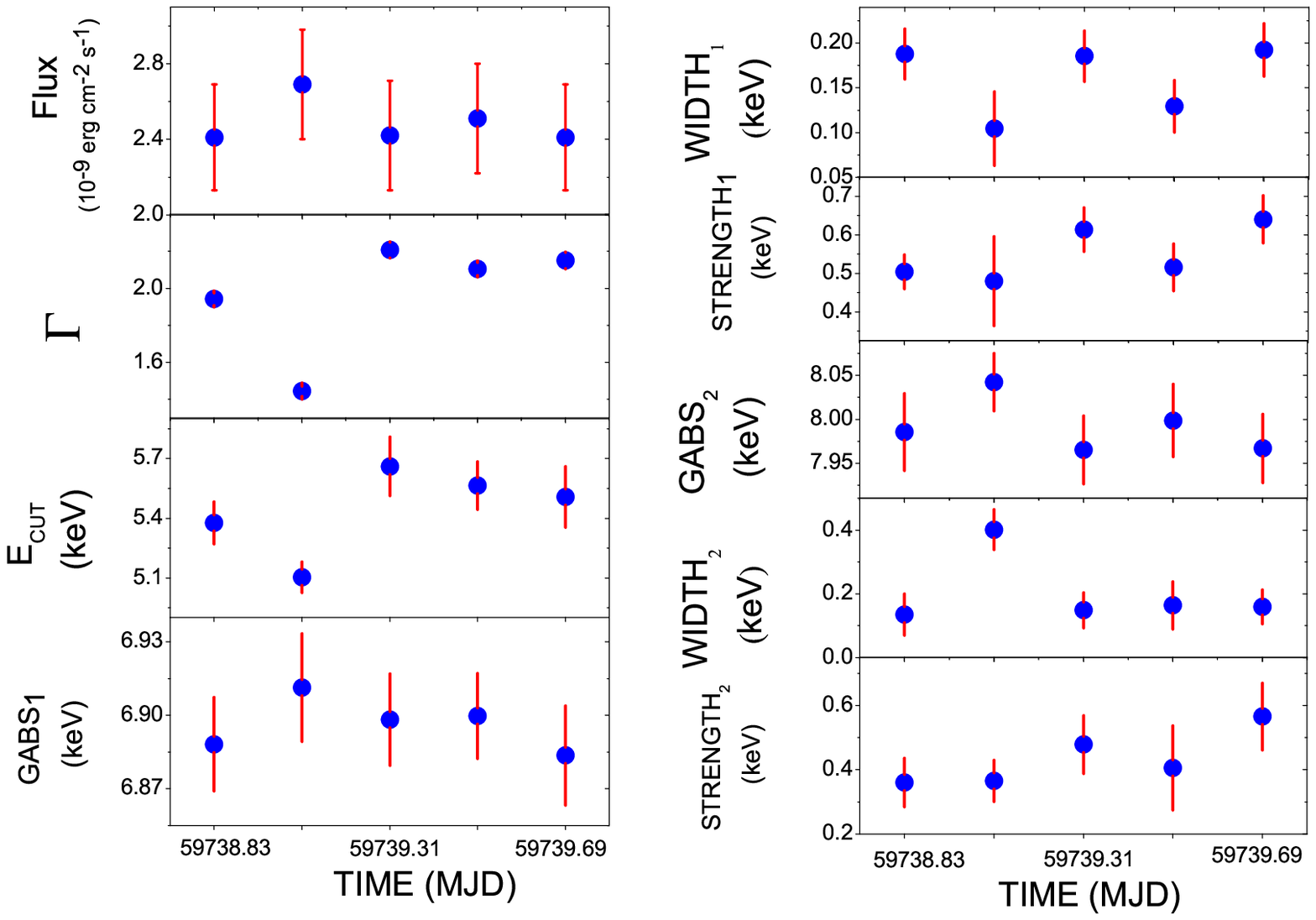}
\end{center}
\caption{Variation of spectral parameters: Flux, Photon Index ($\Gamma$), Cut-off energy ($E_{CUT}$) and absorption line parameters with time in the range (59738.83- 59739.74) MJD using NuSTAR observation in (3-79) keV energy range.}
\label{7}
\end{figure}

In order to investigate the continuum evolution and variabilities in spectral parameters with time and luminosity, the source light curve was divided into five small segments. The NuSTAR light curve of the source with the five segments is presented in Figure \ref{6}. The corresponding spectral products were generated by using a good time interval (GTI) for each segment. The CUTOFFPL model was implemented for spectral fitting of the spectra in the  energy range. The spectra corresponding to all the segments have been well fitted using MODEL I. The variation of the spectral parameters with time has been presented in Figure \ref{7}. The flux has been estimated in the (3-40) keV energy range for each segment. The variation of flux along the segments is presented in the top panel (left) of Figure \ref{7}. The power law photon index exhibits a relative variability throughout the observation, suggesting a harder spectrum along the second segment and a relatively soft spectrum along the third segment. The cut-off energy is also found to vary throughout the observation. Similarly, the parameters corresponding to the absorption features are found to exhibit some complex variabilities throughout the observations. As evident from Figure \ref{7}, the spectral parameters do not follow a sequential increasing or decreasing trend with time. 

\begin{figure}

\begin{center}
\includegraphics[angle=0,scale=0.33]{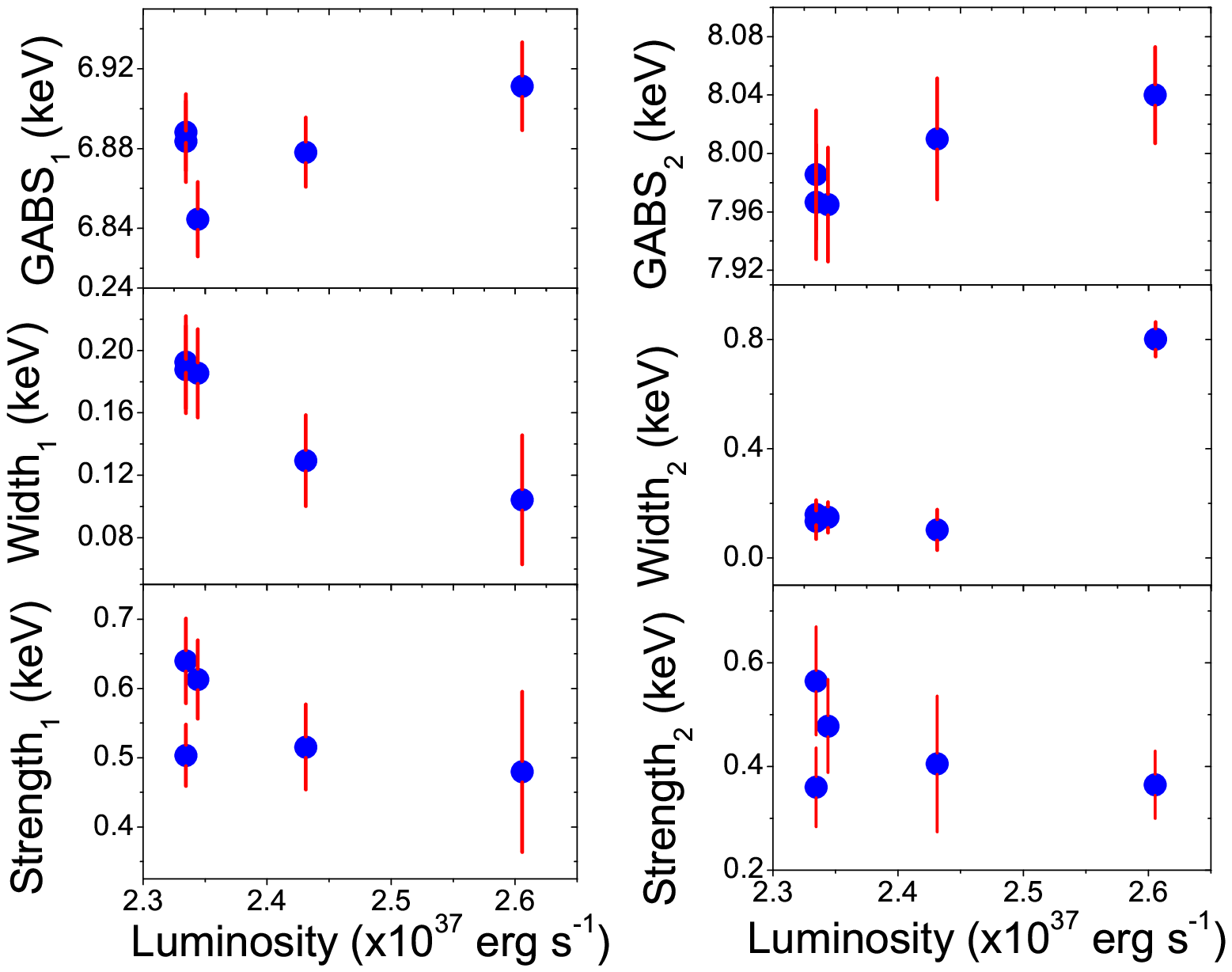}
\end{center}
\caption{Luminosity dependence of absorption line parameters corresponding to NuSTAR observation. The subscripts 1 and 2 denote the K $\alpha$ line parameters and the other absorption line (a blend of Fe XXV and Ni XXVII transitions) parameters
respectively.}
\label{8}
\end{figure}

\begin{table*}
\begin{center}
\begin{tabular}{ccccc}
\hline										
Parameters			&	MODEL I (BB+CUTOFFPL) 	&	&		MODEL II (BB+CompTT)	&	\\
\hline										
$C_{FPMA}$			&	1(fixed)	&	&		1(fixed)	&	\\
$C_{FPMB}$			&	0.983$\pm$0.001	&	&		0.973$\pm$0.002	&	\\
$C_{NICER}$ &	0.912$\pm$0.002	&	&		0.909$\pm$0.004	&	\\
$n_{H}\;(\times 10^{22} cm^{-2})$			&	0.53$\pm$0.02	&	&		0.64$\pm$0.02	&	\\
BB (kT) (keV) &      1.114 $\pm$0.002                        &  & 1.005 $\pm$0.007                 
               &\\
CompTT ($T_{o}$) (keV)			&	-	&	&		0.15$\pm$0.02	&	\\
CompTT (kT) (keV)			&	-	&	&		2.65$\pm$0.02	&	\\
CompTT ($\tau$) 			&	-	&	&		6.02$\pm$0.07	&	\\
$\Gamma$			&	1.141$\pm$0.007&	&			-&	\\
$E_{cut}$ (keV)			&	4.183$\pm$0.003	&	&		-	&	\\
Fe line (keV)			&	6.372$\pm$0.006	&	&			6.34$\pm$0.06&	\\
$\sigma_{Fe}$ (keV)			&	1.014$\pm$0.007	&	&		1.10$\pm$0.04	&	\\
$E_{gabs1}$	(keV)		&	6.92$\pm$0.01	&	&		6.92$\pm$0.02	&	\\
$\sigma_{gabs1}$ (keV)			&	0.121$\pm$0.002	&	&		0.118$\pm$0.002	&	\\
$Strength_{gabs1}$(keV)			&	0.293$\pm$0.002	&	&		0.23$\pm$0.01	&	\\
$E_{gabs2}$	(keV)		&	7.98$\pm$0.02	&	&		7.97$\pm$0.02	&	\\
$\sigma_{gabs2}$ (keV)			&	0.13$\pm$0.02	&	&		0.11$\pm$0.03	&	\\
$Strength_{gabs2}$ (keV)			&	0.36$\pm$0.05	&	&		0.29$\pm$0.01	&	\\
$E_{gabs3}$ (keV)			&	10.24$\pm$0.06	&	&		10.14$\pm$0.03	&	\\
$\sigma_{gabs3}$ (keV)			&	0.94$\pm$0.07	&	&		0.88$\pm$0.04	&	\\
$Strength_{gabs3}$ (keV)			&	2.0(fixed)	&	&		2.0(fixed)	&	\\
$\chi^{2}$ per degrees of freedom			&	2169/1833	&	&	2080/1814		& 	\\										
										
$\chi^{2}_{\nu}$			&	1.18	&	&		1.15	&	\\

 \hline
 \end{tabular}
 \caption{The best fit parameters obtained after joint fitting of NuSTAR/NICER spectra in (0.7-40) keV energy range. The fit statistics $\chi_{\nu}^{2}$  represents reduced $\chi^{2}$ ($\chi^{2}$ per degrees of freedom). Errors quoted for each parameter are within 68\% confidence interval.  The parameter nH represents neutral hydrogen column density, $\Gamma$ represents photon-index, $E_{cut}$ is the cut-off energy, and BB (kT) represents the blackbody temperature. The parameters $T_{o}$, kT, and $\tau$ represent the the soft comptonization temperature, plasma temperature and optical depth corresponding to the CompTT component. The parameters (centroid energy, equivalent width and strength) corresponding to the emission and absorption lines are specified in the table.}
  \end{center}
 \end{table*}

Further, we examined the luminosity dependence of the spectral parameters corresponding to the absorption features. The centroid energy corresponding to the $K_{\alpha}$  absorption line is found to be strongly correlated with luminosity (Pearson's correlation coefficient $\alpha$ = 0.86). However, the equivalent width and strength of the line is found to be anti-correlated with luminosity and is characterized by a correlation coefficient of -0.93 (strong anti-correlation) and -0.4 (weak anti-correlation) respectively. Similarly, centroid energy corresponding to the $K_{\beta}$ absorption line is also found to exhibit a very strong correlation with luminosity ($\alpha$ = 0.97). The equivalent width of the line is also found to exhibit a strong positively correlated ($\alpha$ = 0.96) trend with luminosity while the corresponding strength is found to be weakly anti-correlated ($\alpha$ = -0.5) with luminosity. The equivalent widths of the absorption features are found to vary over the observations revealing variations in the ionization state of the material. The luminosity dependence of the parameters corresponding to the absorption line features has been presented in Figure \ref{8}.
 
 \section{Discussion \& Conclusion}
The available NuSTAR and NICER data have been used to analyze the characteristics of the LMXB source 1A 1744-361. The joint NUSTAR/NICER spectra is fitted in the broadband (0.7-40) keV energy range. The corresponding fit parameters have been presented in Table 3. In addition to the hump observed at around 6.4 keV, denoting the existence of a reflection component by the optically thick accretion disk around the NS-LMXB, multiple absorption features were prominent in the X-ray continuum.  We have fitted the broadband X-ray spectra to reveal the reflection and absorption features and hence determined the model that can suitably describe the continuum. The best-fit parameters corresponding to the continuum models may be interpreted as phenomenological results.

The spectra of such sources often reveal reflection components arising due to Compton scattering of photons leaving the hot corona with the cold electrons in the top layers of the inner accretion disc. The reflection component reveals discrete features because of fluorescent emission and photoelectric absorption by heavy ions in the accretion disc. Emission lines from Fe atoms in the range 6.4 to 6.97 keV are identified with $K_{\alpha}$ radiative transition of iron at various ionization states \citep{39, 40, 41, 42, 43, 44, 45, 46, 47}. Such features are observed in the accretion disc lying closer to the compact object, where rotation of matter is rapid \citep{48, 49}. The reflection spectrum is known to be modified due to transverse Doppler shifts, Doppler broadening, relativistic Doppler boosting, and gravitational redshift, which is responsible for producing characteristic broad and skewed line profile \citep{50}. The emission line is found to be broad in our present study. Such broad iron lines have been observed in many LMXBs, including dippers \citep{55, Ludlam}. Analyzing high-resolution spectra can figure out the source of the broadening by understanding the detailed structure of the lines.  

The centroid energy of the absorption line at 6.92 keV is suggestive of resonant scattering of the $K_{\alpha}$ line from hydrogen-like iron (e.g., \cite{ Neilsen, Honghui Liu et al 2022 ApJ 933 122, Ueda}). The centroid energies corresponding to Ni $K_{\alpha}$ and Ni $K_{\beta}$ lines are 7.5 keV and 8.26 keV respectively \citep{Sako, Palmeri}. Only a few astrophysical X-ray spectra have identified prominent nickel line detections. This may be caused by a combination of weak
spectral resolution in this band, which makes it difficult to distinguish between nickel $K_{\alpha}$ and iron $K_{\beta}$
lines, and inadequate counting statistics at energies
above 7 keV, where the nickel lines occur \citep{Palmeri}. Therefore, the absorption feature observed at 7.98 keV may be attributed to a blend of Fe XXV and Ni XXVII transitions \citep{Neilsen}. Also, the hypothesis that such absorption features are unique to superluminal jet sources containing a black hole can be ruled out indicating that common mechanisms are responsible for producing absorption lines and are independent of the nature of the compact object \citep{Uedaa, Ueda, Yamaoka, Kotani, Lee}. Furthermore, the presence of the absorption line in the X-ray spectrum indi-
cates the existence of a highly ionized gas along the
line of sight. This gas is not in a spherically symmetric configuration, as the re-emitted lines produced in other directions would fill up the absorption-line feature, rendering it undetectable \citep{Ueda}. Such a physical condition leading to iron K absorption lines is commonly observed in many X-ray binaries accreting  via disks, due to the occurrence of a moderately thick accretion-disk flow in a high-ionization state. The parameters corresponding to the absorption lines provide valuable information about the physical condition of the line-absorbing matter. The centroid energy is related to the ionization state of the hot plasma, while the equivalent width is connected to the column density of the plasma. The absorbed flux of the source in (0.7-40) keV energy range was found to be $\sim3.71 \;\times\;10^{-9}\;erg\;cm^{-2}\;s^{-1}$ resembling a luminosity of $\sim3.59 \;\times\;10^{37}\;erg\;s^{-1}$ assuming a distance of 9 kpc.

As evident from the HID of the atoll source 1A 1744-361, high intensities correspond to the typical flat hardness of the banana state , and low-intensity data indicate the hardening of the island state. During our observations, the source is found to be in the banana state of the HID. The mass accretion rate or luminosity determines the position on the banana track. Similarly, the CCD reflects that the soft-state observations correlate to the banana state, whereas the hard-state observations represent the island state. There exists a thermal equilibrium between the neutron star and the extended accretion
disc corona (ADC) in all the LMXB sources with luminosities greater than (1-2) $\times 10^{37} erg\; s^{-1}$ \citep{Church}. In the context of our current study, this condition is satisfied. However, in atoll sources that are less luminous than (1-2) $\times 10^{37} erg\; s^{-1}$ are often identified with the island state. Thermal equilibrium collapses in the island state, causing the ADC temperature to increase above the blackbody temperature of the neutron star.

The continuum evolution and variabilities in spectral parameters with time have been observed using time-resolved spectroscopy by dividing the source light curve into five small segments. Most of the spectral parameters were found to vary moderately with time. The parameters corresponding to the absorption features were found to exhibit a significant luminosity dependence. The centroid energy of the $K_{\alpha}$ absorption line is found to be strongly correlated with luminosity while its equivalent width is found to exhibit a strong anti-correlation with luminosity. Similarly, the centroid energy and the equivalent width of the second absorption line is found to be strongly correlated with luminosity. However, the strength of these absorption lines is found to exhibit a weak anti-correlation with luminosity.

\section*{Acknowledgements}
This work has been performed by using publicly available data of the source provided by NASA HEASARC data archive. We are grateful to IUCAA Centre for Astronomy Research and Development (ICARD), Department of Physics, NBU, for providing research facilities. We would like to express our gratitude to the anonymous reviewer for the comments and suggestions.

\section*{Data availability}
 
The observational data used in this research work can be accessed from the HEASARC data archive and is publicly available.

\bsp	
\label{lastpage}
\end{document}